\documentclass[twocolumn,preprintnumbers,amsmath,amssymb]{revtex4}
\usepackage{graphicx} 
\usepackage{setspace} 
\usepackage{dcolumn} 
\usepackage{bm} 
\usepackage{natbib}
\usepackage{adjustbox}
\usepackage{hyperref}
\hypersetup{colorlinks=true,linkcolor=blue,citecolor=blue,urlcolor=blue}
\usepackage{epsfig}

\begin{document}
\title{Thermoelectric properties of Sb doped AlFe$_2$B$_2$}
\author{Duraisamy Sivaprahasam$^{1,}$\footnote{Email: sprakash@arci.res.in}}
\author{Ashutosh Kumar$^{2}$}
\author{Babu Jayachandran$^1$}
\author{Raghavan Gopalan$^{1}$}

\affiliation{$^1$Centre for Automotive Energy Materials (CAEM), International Advanced Research Centre for Powder Metallurgy and New Materials (ARCI), IIT Madras Research Park, Taramani, Chennai – 600 113, INDIA}
\affiliation{$^2$ICMMO (UMR CNRS 8182), Université Paris-Saclay, F-91405 Orsay, France}

\date{\today}
\begin{abstract}
In this work, thermoelectric properties of Al$_{1.2}$Fe$_2$B$_2$ compound were investigated over a temperature range from 300\,K to 773\,K. Al$_{1.2}$Fe$_2$B$_2$ compound was produced by vacuum arc melting of Al, Fe, and B followed by annealing at 1323 K under argon atmosphere. The annealed ingots were subsequently crushed into powder and hot pressed at 1273 K under vacuum. The hot-pressed alloy predominantly contained Al$_{1.2}$Fe$_2$B$_2$ phase with a small fraction of FeB, which decreases further upon 0.1 \% Sb doping in Al$_{1.2}$Fe$_2$B$_2$. The pristine Al$_{1.2}$Fe$_2$B$_2$ exhibits n-type conductivity with a maximum figure of merit (zT) of 0.03 at 773\,K. The Sb doping improves the Seebeck coefficient at high temperatures and also reduces the phonon thermal conductivity across the temperature range studied. The decrease in phonon thermal conductivity is attributed to the point-defect phonon scattering due to mass fluctuation between the Fe and Sb atoms. The 0.1 at\% Sb doping at the Fe site results in improved zT of 0.056 at 773\,K in spite of its limited dissolution in Al$_{1.2}$Fe$_2$B$_2$ and forms FeSb$_2$ secondary phase.
\end{abstract}
\maketitle
\section{Introduction}
Thermoelectric (TE) power generation in recent years has been explored for several innovative applications\cite{R1}. The direct conversion of heat into electricity using this solid–state method
provides an opportunity to produce power from different thermal energy sources. The utility of a TE device depends on TE material properties, which are quantified by a figure of merit (zT), defined as zT = ($\alpha^2\sigma$T)/$\kappa$. An optimized value of the Seebeck coefficient ($\alpha$), electrical conductivity ($\sigma$) and thermal conductivity ($\kappa$) is desired to obtain enhanced zT, owing to the interdependence of these parameters on carrier concentration, band structure, and charge scattering\cite{R2}. Promising p– and n-type TE materials such as PbTe, GeTe, and CoSb$_3$ have been investigated following several novel approaches including band convergence\cite{R3}, Fermi-level optimization \cite{R4}, resonance level doping \cite{R5}, crystal structure modification \cite{R6, R7} high-entropy concept\cite{R8, R8a}, composite methods \cite{R9, R9a} etc. However, the relatively low TE conversion efficiency and high material costs are the major challenges that inhibit large-scale applications of these materials. Replacing the expensive compounds with environment-friendly and abundantly available elements can significantly reduce device costs.\\
Al$_{1.2}$Fe$_2$B$_2$ is one such compound; however, not been investigated in detail so far for its TE properties. It is a ferromagnetic compound known for its Curie temperature close to room temperature with promising magnetocaloric applications.\cite{R10, R11} Several methods, including melting \cite{R10, R11}, melt spinning \cite{R12}, and spark plasma sintering of elemental powders \cite{R13} have been used to synthesize this compound. The phase formation from the melt during solidification follows peritectic reaction and demands proper control of Al content to avoid the impurity phases viz. FeB and Al$_{13}$Fe$_4$ \cite{R10, R11, R12}. The FeB forms when Al content is sub-stoichiometry, whereas the excess amount leads to the formation of Al$_{13}$Fe$_4$. The Al$_{13}$Fe$_4$ impurity is usually removed by etching with dilute HCl acid. \cite{R10, R11, R12}\\
Al$_{1.2}$Fe$_2$B$_2$ crystallizes in a layered orthorhombic structure in which, Fe$_2$B$_2$ slabs are stacked between aluminum layers.\cite{R14} It shows metallic nature that results in higher $\sigma$ and $\kappa$, and, lower $\alpha$, and hence, requires optimization of electronic structure to improve TE properties. With this aim, we demonstrate the effect of Sb alloying on TE properties of earth-abundant Al$_{1.2}$Fe$_2$B$_2$. The Sb substitution has shown improved TE properties, i.e. increased Seebeck coefficient and electrical resistivity compared to the parent compound.\\
\section{Experimental Details}
\begin{figure*}
\centering
\includegraphics[width=0.65\linewidth]{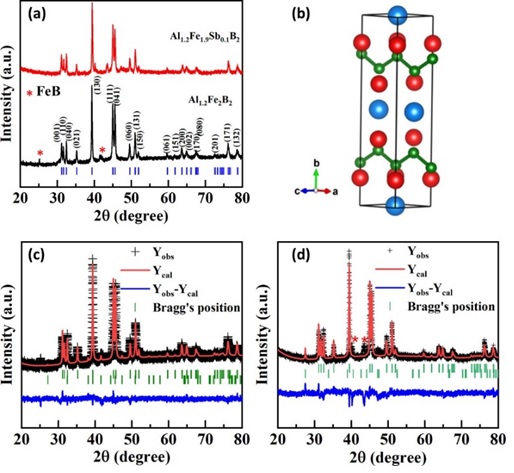}
\caption{(a) XRD patterns of Al$_{1.2}$Fe$_2$B$_2$ and Sb doped Al$_{1.2}$Fe$_2$B$_2$, (b) structure of Al$_{1.2}$Fe$_2$B$_2$ obtained from vesta software using Rietveld refinement parameters. Rietveld refinement pattern for (c) Al$_{1.2}$Fe$_2$B$_2$, and (d) Sb doped Al$_{1.2}$Fe$_2$B$_2$.}
\end{figure*}
The Al$_{1.2}$Fe$_2$B$_2$ and Al$_{1.2}$Fe$_{1.9}$Sb$_{0.1}$B$_2$ compounds were synthesized by vacuum arc melting of Al
(99.9\%, American elements), Fe (99.9\%, 3.2–6.4\,mm, Alfa Aesar), B (99.5\%, 20 mm, Thermofisher Scientific) and Sb (99.9\% American elements) were weighed according to the stoichiometric composition. A 20 wt.\% excess Al was added to compensate for the losses during arc melting. The solidified ingots were annealed for 96\,h at 1323\,K under argon, and ground into powder. The powder was subsequently consolidated by a vacuum (5$\times$10$^{-5}$ mbar) hot pressing at 1273\,K for 2\,h under 50 MPa pressure to 20\,mm diameter pellets. The structural characterization was performed using a Rigaku Smart Lab X–ray powder diffractometer (Cu-K$_{\alpha}$, $\lambda$=0.15406\,nm) with a step of 0.01 and scan range from 20–80$^{\circ}$ at 1$^\circ$/min scan rate. The microstructural observation was done using field emission scanning electron microscope (FE–SEM; Zeiss Merlin, Germany). Chemical analysis of various phases present in the microstructure was carried out using energy dispersive X–ray spectrometry (EDS). The Seebeck coefficient, electrical conductivity was measured using a standard four-probe method
(Seebsys system: NorECS AS, Norway). Thermal conductivity ($\kappa$=D$\times$ C$_p$ $\times\rho$) was measured using the thermal diffusivity (D), specific heat (C$_p$) and sample density ($\rho$). The D
measurements were performed using the laser flash technique (LFA–457, Netsch GmbH, Germany) on a cylindrical sample having 10 mm diameter and $\sim$2 mm thickness. The C$_p$ was obtained using Dulong–Petit law. The sample density ($\rho$) was measured using the Archimedes method.\\
\begin{table}[h]
\caption{Rietveld refinement parameters obtained from the two-phase (Al$_{1.2}$Fe$_2$B$_2$, s.g.:
\textit{Pbnm} is the primary phase and FeB, s.g.:\textit{Pbnm} is the second phase) Rietveld
refinement of the XRD pattern of Al$_{1.2}$Fe$_2$B$_2$ and Al$_{1.2}$Fe$_{1.9}$Sb$_{0.1}$B$_2$}
\centering
\begin{adjustbox}{width=0.5\textwidth}
    \small
\begin{tabular}{c c c c c c c}
\hline
sample & Phases  & Lattice parameters & & & Phase  & $\chi^2$ \\
        &         & a ($\AA$) & b ($\AA$) & c ($\AA$) &\% & \\ 
\hline
\hline
Al$_{1.2}$Fe$_{2}$B$_2$ & Al$_{1.2}$Fe$_{2}$B$_2$ & 2.924 (5) & 11.033 (9) &2.871 (1) &93.12&1.42\\
            & FeB & 4.095 (5) & 5.520 (6) & 2.886 (8) & 6.88\\
            
Al$_{1.2}$Fe$_{1.9}$Sb$_{0.1}$B$_2$ & Al$_{1.2}$Fe$_{1.9}$Sb$_{0.1}$B$_2$ & 2.915 (4) & 11.025 (8) & 2.873 (7) & 99.42 & 2.25\\
 & FeB & 4.102 (5) & 5.306 (4) & 2.803 (8) & 0.58\\

\hline
\end{tabular}
  \end{adjustbox} 
\label{Table I}
\end{table}
\section{Results and Discussion}
\begin{figure*}
\centering
\includegraphics[width=0.99\linewidth]{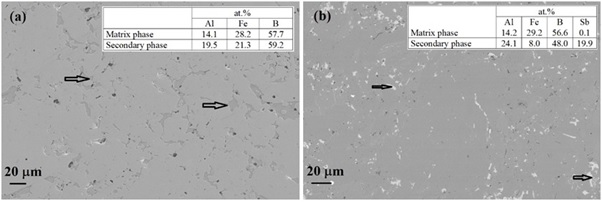}
\caption{FE–SEM images of hot pressed (a) Al$_{1.2}$Fe$_2$B$_2$ and (b) Al$_{1.2}$Fe$_{1.9}$Sb$_{0.1}$B$_2$ samples.}
\end{figure*}
Fig. 1(a) shows the XRD patterns of hot-pressed Al$_{1.2}$Fe$_2$B$_2$ and Al$_{1.2}$Fe$_{1.9}$Sb$_{0.1}$B$_2$ samples. The major peaks in the pattern can be indexed to Al$_{1.2}$Fe$_2$B$_2$ [ICSD reference: 98–000–7593] having
an orthorhombic crystal structure, with a few minor peaks corresponding to FeB [ICSD reference:
00-002-0869] impurity (marked by $*$ in Fig.~1(a)). It is found that a peak corresponding to
Al$_{13}$Fe$_4$ phase is seen in the Al$_{1.2}$Fe$_2$B$_2$ synthesized with 30\% excess Al samples (XRD pattern
not given). It indicates that 20\% excess Al addition in the starting stoichiometry compensates
for the losses during synthesis by arc melting, which is optimum. Further, the XRD diffraction
pattern of the Sb substituted sample shows a similar diffraction pattern, with additional two minor peaks
of FeSb$_2$ are also observed at 40.2$^\circ$ and 44.1$^\circ$, which indicates that 0.1 at\% of Sb is the higher side of the solubility limit. The intensity corresponding to the FeB phase in Sb-doped Al$_{1.2}$Fe$_2$B$_2$ sample reduces, indicating the Sb doping helps to obtain a near single phase of Al$_{1.2}$Fe$_2$B$_2$. Further, two-phase Rietveld refinement of the XRD pattern was done using Fullprof software, considering Al$_{1.2}$Fe$_2$B$_2$ as primary (\textit{Pbnm}) and FeB (\textit{Pbnm}) as the secondary phases. The refinement patterns for Al$_{1.2}$Fe$_2$B$_2$ and Sb-doped Al$_{1.2}$Fe$_2$B$_2$ are shown in Fig.~1(c-d) and the corresponding refinement parameters are shown in Table~I. It is seen that the phase fraction for the FeB, obtained from the refinement, reduces in Sb doped sample. The goodness of fit ($\chi^2$) is in the acceptable limit for both samples. Further, the refinement parameters are used to create the structure of the unit cell using Vesta software\cite{R14a} as shown in Fig.~1b.\\ 
\begin{figure*}
\centering
\includegraphics[width=0.65\linewidth]{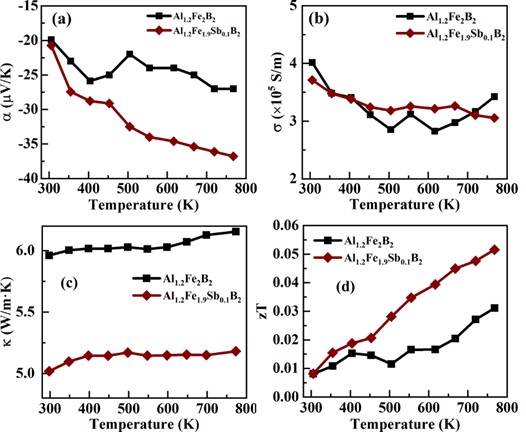}
\caption{Temperature dependent (a) Seebeck coefficient ($\alpha$), (b) electrical conductivity
($\sigma$), (c) thermal conductivity ($\kappa$) and (d) figure of merit (zT) for both Al$_{1.2}$Fe$_2$B$_2$ and Al$_{1.2}$Fe$_{1.9}$Sb$_{0.1}$B$_2$ compounds.}
\end{figure*}
Fig. 2(a–b) shows the SEM micrographs of Al$_{1.2}$Fe$_2$B$_2$ and Sb doped Al$_{1.2}$Fe$_2$B$_2$ samples. The
microstructure of Al$_{1.2}$Fe$_2$B$_2$ is homogeneous with a small volume of secondary phase/s constituted
of Fe, Al, and B. The atomic ratio values mentioned in the inset of Fig.~2(a,b) are the average 
values taken at various spots within the cross-section of the sample. As it is difficult to quantify
boron using EDS, Al: Fe ratio was used to measure the stoichiometry. The ratio in Al$_{1.2}$Fe$_2$B$_2$
sample is 0.46 for region AlFe$_{1+x}$B$_2$, which is close to the ratio for Al$_{1.2}$Fe$_2$B$_2$ compound (0.5)
and the secondary phase/s is the (FeAl)B$_{1+x}$ phase. From the XRD results, it is evident that the
secondary phase/s is predominantly constituted of FeB, however, the EDS analysis suggests the
composition of the second phase region is a nearly equiatomic ratio of Al and Fe. From the
established Al-Fe-B ternary equilibrium phase diagram\cite{R15, R16}, it is clear that the line
compound Al$_{1.2}$Fe$_2$B$_2$, under off stoichiometry of Al depletion, form FeB, FeAl$_2$ and B phases
along with Al$_{1.2}$Fe$_2$B$_2$ phase. The comparable atomic fraction observed in the EDS analysis for
both Fe and Al suggests that the secondary phases observed may also contain the FeAl$_2$ phase apart
from FeB. In Al$_{1.2}$Fe$_{1.9}$Sb$_{0.1}$B$_2$ the Al to Fe ratio was 0.48–0.49, and the secondary phases
consisted of relatively higher Sb than matrix Al$_{1.2}$Fe$_2$B$_2$ phase. The 0.1 at.\% Sb added largely
makes into secondary phase/s, and only a small fraction dissolves in Al$_{1.2}$Fe$_2$B$_2$ matrix. The
presence of minor peaks in the XRD pattern of Sb doped Al$_{1.2}$Fe$_2$B$_2$Sb closely matches with
FeSb$_2$ phase strengthening the view that Sb mostly segregates to the secondary phase. AlB$_2$ and
AlB$_{12}$ are some of the other possible constituents that can form as secondary phases under off stoichiometry of Fe depletion from Al$_{1.2}$Fe$_2$B$_2$ stoichiometry. However, no observable peak
corresponding to these phases was present in the XRD pattern.\\
Fig.~3(a) depicts the temperature-dependent Seebeck coefficient ($\alpha$) for Al$_{1.2}$Fe$_2$B$_2$ and
Al$_{1.2}$Fe$_{1.9}$Sb$_{0.1}$B$_2$ samples. The negative values for $\alpha$ indicate that the samples have dominant n-type carriers. The value of $\alpha$ ($\sim$-20 $\mu$V/K at 300\,K) obtained in pristine Al$_{1.2}$Fe$_2$B$_2$ matches with the reported values in the literature.\cite{R18} The Sb substitution improves $\alpha$; however, its effect is particularly notable at elevated temperatures. The $\alpha$ for Al$_{1.2}$Fe$_{1.9}$Sb$_{0.1}$B$_2$ is $\sim$-38 $\mu$V/K at 773\,K higher than that of pristine Al$_{1.2}$Fe$_{1.2}$B$_2$ ($\sim$-27 $\mu$V/K). The electrical conductivity ($\sigma$) as a function of temperature is shown in Fig.~3(b). The measured $\sigma$ ($\sim$ 4$\times$10$^5$ S/m at 300\,K) is slightly higher than the value reported (3.7$\times$10$^5$ S/m at 300\,K) for Al$_{1.2}$Fe$_2$B$_2$ compound.\cite{R17} This difference in $\sigma$ can be attributed to the chemical homogeneity
in the present work, as the samples were prepared using arc melting, annealing, crushing, and
hot pressing. Whereas, the samples reported in the literature were arc-melted followed by
annealing. The Sb alloying seems to be marginally affecting the $\sigma$. However, at elevated
temperatures, the electrical conductivity of pure and Sb-doped samples is comparable. Fig.~3(c) shows the temperature-dependent thermal conductivity ($\kappa$) for Al$_{1.2}$Fe$_2$B$_2$ and
Al$_{1.2}$Fe$_{1.9}$Sb$_{0.1}$B$_2$. The thermal conductivity for Al$_{1.2}$Fe$_2$B$_2$ is $\sim$6 W/m·K at 300\,K, which is in good agreement with the previously reported value.\cite{R18} The Sb doping noticeably decreases the thermal conductivity across the temperature range of testing. It is known that the total thermal conductivity consists of electronic ($\kappa_e$) as well as phonon thermal conductivities ($\kappa_{ph}$) i.e.,($\kappa=\kappa_e+\kappa_{ph}$). The $\kappa_e$ is proportional to the electrical conductivity, according to Wiedemann Franz law. From Fig.~3(b), it was also found that the electrical conductivity was almost similar for both the samples, and hence their electronic contribution to thermal conductivity might be similar, considering the metallic nature of the sample. This further suggests that Sb doping reduces the phonon thermal conductivity resulting in lower thermal conductivity. This decrease in lattice thermal conductivity may be attributed to the point-defect phonon scattering by mass fluctuation between host atoms (Fe) and dopants (Sb).\cite{R19} Further, as found in the XRD and SEM analysis, some FeSb$_2$ impurity present in the Sb-doped Al$_{1.2}$Fe$_2$B$_2$ sample further result in reduced thermal conductivity.

The thermoelectric parameters ($\alpha$, $\sigma$, and $\kappa$) measured in this study have been used to calculate the thermoelectric figure of merit (zT) and its variation with temperature is shown in Fig.~3(d). The zT for pristine Al$_{1.2}$Fe$_2$B$_2$ sample is 0.03 at 773\,K, which is almost the first report in t conductivity in Sb doped Al$_{1.2}$Fe$_2$B$_2$ results in improved zT. A maximum zT of 0.056 is achieved at 773\,K for Al$_{1.2}$Fe$_{1.9}$Sb$_{0.1}$B$_2$. Moreover, the compound having earth-abundant elements with enhanced electrical conductivity and the possibility to reduce the phonon thermal
conductivity makes it interesting for further investigation concerning the optimization of carrier
concentration and band structure, which can significantly improve its TE performance.\\

\section{Conclusion}
Earth-abundant Al$_{1.2}$Fe$_2$B$_2$ and Al$_{1.2}$Fe$_{1.9}$Sb$_{0.1}$B$_2$ are synthesized using a vacuum arc melting. The X-ray diffraction analysis followed by two-phase Rietveld refinement showed
the presence of a smaller quantity (6.9 wt.\%) of FeB secondary phase along with Al$_{1.2}$Fe$_2$B$_2$. The
phase fraction corresponding to the FeB phase reduces (0.6 wt.\%) in Sb doped Al$_{1.2}$Fe$_2$B$_2$ sample. The metallic nature of pristine Al$_{1.2}$Fe$_2$B$_2$ shows a high electrical conductivity but a low Seebeck coefficient and large thermal conductivity. However, Sb doping improves the Seebeck
coefficient with almost similar electrical conductivity and decreases the total thermal
conductivity via decreased phonon thermal conductivity. As a result of the simultaneous
increase in the Seebeck coefficient and decrease in thermal conductivity, an improved figure of
merit (zT) of 0.056 at 773 K for Al$_{1.2}$Fe$_{1.9}$Sb$_{0.1}$B$_2$. 

\newpage
\end{document}